\documentclass[twocolumn,aps,superscriptaddress,showpacs,nofootinbib,floatfix,preprintnumbers,amsmath,amssymb]{revtex4}

\usepackage{epsfig,bm}

\usepackage{graphics}

\usepackage[normalem]{ulem}  % \sout{old text} for strikeout
\usepackage[dvips]{color} % For blue in-text comments and additions

\renewcommand\sout{\bgroup \color{red} \ULdepth=-.5ex \ULset}

\begin{document}

%%%%%%%%%%%%%%%%%%%%%%%%%%%%%%%%%%%%%%%%%%%%%%%%%%%%%%%%%%%%%%%%%%%%%%%%%%
\title{Quantifying jet transport properties via large $p_T$ hadron production}
%%%%%%%%%%%%%%%%%%%%%%%%%%%%%%%%%%%%%%%%%%%%%%%%%%%%%%%%%%%%%%%%%%%%%%%%%%

\author{Zhi-Quan Liu, Hanzhong Zhang, Ben-Wei Zhang and Enke Wang}

\affiliation{Institute of Particle Physics and Key Laboratory of Quark $\&$ Lepton Physics (MOE), Central China Normal University, Wuhan 430079,
China}

%\vspace{1.4in}

\begin{abstract}

Nuclear modification factor $R_{AA}$ for large $p_T$ single hadron is studied
in a next-to-leading order (NLO) perturbative QCD (pQCD) parton model
with medium-modified fragmentation functions (mFFs) due to jet quenching
in high-energy heavy-ion collisions. The energy loss of the hard partons in the QGP is incorporated in the mFFs which utilize two most
important parameters to characterize the transport properties of the hard parton jets: the jet transport parameter $\hat q_{0}$ and the mean
free path $\lambda_{0}$, both at the initial time $\tau_0$.
A phenomenological study of the experimental data for $R_{AA}(p_{T})$ is performed to
constrain the two parameters with simultaneous $\chi^2/{\rm d.o.f}$ fits to RHIC as well as LHC data.
We obtain for energetic quarks $\hat q_{0}\approx 1.1 \pm 0.2$ GeV$^2$/fm and $\lambda_{0}\approx 0.4 \pm 0.03$ fm in central $Au+Au$
collisions at $\sqrt{s_{NN}}=200$ GeV, while $\hat q_{0}\approx 1.7 \pm 0.3$ GeV$^2$/fm, and $\lambda_{0}\approx 0.5 \pm 0.05$ fm in central
$Pb+Pb$ collisions at $\sqrt{s_{NN}}=2.76$ TeV.
Numerical analysis shows that the best fit favors a multiple scattering picture for the energetic jets propagating through the bulk medium,
with a moderate averaged number of gluon emissions. Based on the best constraints for $\lambda_{0}$ and $\tau_0$, the estimated value for the
mean-squared transverse momentum broadening is moderate which implies that the hard jets go through the medium with small reflection.

\vspace{12pt}

%\noindent {\em PACS numbers:} 12.38.Mh, 24.85.+p, 25.75.-q
\end{abstract}


\maketitle

\section{Introduction}

A strongly coupled quark gluon plasma (sQGP) consisting of deconfined quarks and gluons may have been created in the central region of
high-energy
nuclear collisions at the BNL Relativistic Heavy Ion Collider (RHIC) and the CERN Large Hadron Collider (LHC). One important evidence for the
formation of sQGP from the experiment results are the jet quenching phenomena~\cite{xnw92,Gyulassy94}
that include the strong suppression of single hadron spectra \cite{star-suppression,phenix-suppression,CMS:2012aa, Abelev:2012hxa}, back-to-back
dihadron \cite{star-dihadron,Aamodt:2011vg} and photon-triggered hadron \cite{Adare:2009vd,Abelev:2009gu} as well as jet productions
\cite{Aad:2010bu,Chatrchyan:2011sx} with large
transverse momentum in central $A+A$ collisions as compared to $p+p$ collisions. These observed jet quenching patterns in heavy-ion collisions
at RHIC/LHC can be described well by
different theoretical models~\cite{Vitev:2002pf,Wang:2003mm,
Eskola:2004cr,renk2006,Zhang:2007ja,Qin:2007rn,Zhang:2009rn,Chen:2011vt,Majumder:2011uk,Zapp:2012ak,He:2011pd,Dai:2012am}
that incorporate parton energy loss induced by multiple parton scattering and gluon bremsstrahlung as it propagates
through the dense matter.

A widely used parameter controlling the parton energy loss is the jet transport parameter $\hat q$ \cite{Baier:1996sk},
or the mean-squared transverse momentum broadening per unit
length for a jet in a strong interacting medium, which is also related
to the gluon distribution density of the medium\cite{Baier:1996sk,CasalderreySolana:2007sw} and therefore
characterizes the medium property as probed by an energetic jet.
To estimate the jet transport parameter $\hat q$ intense theoretical studies have been made, such as with the weakly-coupled
QCD~\cite{Kang:2013raa,Wu:2014nca,Ghiglieri:2015zma}, the strong-coupled AdS/CFT~\cite{Liu:2006ug,Zhang:2012jd}, and the lattice
calculations~\cite{Majumder:2012sh,Panero:2013pla}.
Recently, a phenomenological investigation had been carried out to extract the initial value of jet
transport coefficient $\hat q_{0}$ at initial time $\tau_0$, which gives  $\hat q_{0}\approx 1.2
\pm 0.3$ GeV$^2$/fm in Au+Au collisions at $\sqrt{s_{NN}}=200$ GeV and $\hat q_{0}\approx 1.9 \pm 0.7 $ GeV$^2$/fm in Pb+Pb collisions at
$\sqrt{s_{NN}}=2.76 $ TeV for a given quark with initial energy of 10 GeV \cite{Burke:2013yra}.

In this paper we will extract the initial jet transport parameter $\hat q_{0}$ and the initial mean free path $\lambda_{0}$ at initial time
$\tau_0$ on the bulk medium evolution by comparing the experimental data  at RHIC/LHC with numerical simulations of of single hadron yields with
large $p_T$ in a next-to-leading order (NLO) pQCD parton model, where the EPS09 parametrization set of NLO nuclear parton distribution functions
(nPDFs)
has been used to take into account of possible initial-state cold nuclear matter effects, and a phenomenological model
\cite{Zhang:2007ja,Zhang:2009rn} for
medium-modified fragmentation functions calculated in leading-order (LO) at twist-4 in the high-twist approach of jet quenching
\cite{Wang:2001ifa,Zhang:2003yn,Majumder:2007hx} has been
utilized to incorporate parton energy loss. The evolution of bulk medium
used in the study for parton propagation was given by a 3 + 1 dimensional ideal hydrodynamic model \cite{Hirano2001,HT2002} which is constrained
by experimental data on hadron spectra. From calculations with the two independent inputs for the parameters and simultaneous $\chi^2/{\rm
d.o.f}$ fits to the RHIC and the LHC data, we obtain that for energetic quarks $\hat q_{0}\approx 1.1 \pm 0.2$ GeV$^2$/fm and
$\lambda_{0}\approx
0.4 \pm 0.03$ fm in central $Au+Au$ collisions at $\sqrt{s_{NN}}=200$ GeV, while $\hat q_{0}\approx 1.7 \pm 0.3$ GeV$^2$/fm, and
$\lambda_{0}\approx 0.5 \pm 0.05$ fm in central $Pb+Pb$ collisions at $\sqrt{s_{NN}}=2.76$ TeV. This simultaneous and separate constraint of the
two initial values should give a precise and quantitative description for jet quenching to probe the medium properties. For a parton jet
propagating through the bulk medium,
the average transverse momentum broadening squared $\langle q_T^2 \rangle$ depends on the transport parameter as well as the mean free path,
$\langle q_T^2 \rangle = \hat q \lambda$. Our numerical results show that the mean transverse momentum broadening squared of energetic
partons for one scattering at initial time $\tau_0$ in the center of the fireball at LHC is about 2 times of that at RHIC.

The rest of the paper is organized as follows. We first give a brief overview of the NLO pQCD
parton model for single inclusive hadron spectra and a phenomenological model for medium-modified fragmentation functions in
Sec.~\ref{sec:pQCD}. Then the numerical calculations for phenomenological studies of the experimental data on single hadron suppression and
extraction of the jet transport parameter and the mean free path are carried out in Sec .~\ref{sec:result}. We present some discussions in Sec.
\ref{sec:discussion} and finally summarize our study in Sec. \ref{sec:summary}.

\section{NLO pQCD Parton model and modified fragmentation functions}\label{sec:pQCD}

We will utilize the pQCD parton model at NLO for the initial
jet production spectra which has been applied to large
$p_T$ hadron production in high energy hadron-hadron
reactions with great successes~\cite{owens}. In the model the differential cross section of hadron yields has been expressed as a
convolution of NLO parton-parton scattering cross sections, parton
distribution functions (PDFs) in nucleons and parton fragmentation
functions (FFs),
\begin{eqnarray}
\frac{d\sigma_{pp}}{dyd^2p_T}&=&\sum_{abcd}\int dx_adx_b
f_{a}(x_a,\mu^2) f_{b}(x_b,\mu^2) \nonumber \\
&&\hspace{-0.1in}\times \frac{d\sigma}{d\hat{t}}(ab\rightarrow cd)
\frac{D_{h/c}(z_{c},\mu^2)}{\pi z_{c}} +\mathcal {O}(\alpha_s^3) ,
\label{eq:pp}
\end{eqnarray}
where $d\sigma(ab\rightarrow cd)/d\hat{t}$ denotes the leading-order (LO) elementary parton
scattering cross sections at $\alpha_s^2$. The NLO contributions in $\mathcal {O}(\alpha_s^3)$ involve both
$2\rightarrow 3$ tree level processes and one loop virtual corrections
to $2\rightarrow2$ tree processes. Processes at $2\rightarrow 3$ tree level inlcude $qq \rightarrow qqg$, $q\bar q \rightarrow q\bar qg$, $q\bar
q \rightarrow ggg$, $qg \rightarrow qgg$, $qg \rightarrow qq\bar q$, $gg \rightarrow q\bar qg$, $gg \rightarrow ggg$, etc, which include soft
and collinear contributions. A standard $\overline {MS}$ renormalization scheme is applied to control ultraviolet divergence in one loop virtual
corrections
to $2\rightarrow2$ tree processes. More detailed discussions on calculations at NLO could be found in ~\cite{Soper}.
In this paper the numerical calculations are carried
out with a NLO Monte Carlo program~\cite{owens} where two cut-off parameters, $\delta_s$ and $\delta_c$,
are employed to isolate the collinear and soft divergences in the squared
matrix elements of the $2\rightarrow 3$ processes. The regions with
the divergences are integrated over in n-dimension phase space and the results are
added with the squared matrix elements of the $2\rightarrow 2$
processes. This gives a set of two-body and three-body weights depending on
$\delta_s$ and $\delta_c$. But the dependence will be eliminated
after the weights are combined in the calculation of physical observables, and the final numerical results are
insensitive to the cut-off parameters~\cite{owens}.

We employ the same factorized form for the inclusive large $p_T$
particle production cross section in nucleus-nucleus collisions, which
can be computed as a convolution of nuclear thickness functions, the nuclear parton distribution functions (nPDFs),
elementary parton-parton scattering cross sections
and effective medium-modified parton fragmentation functions (mFFs)~\cite{Zhang:2007ja,Zhang:2009rn},
\begin{eqnarray}
\frac{dN_{AB}}{dyd^2p_T}&=&\sum_{abcd}\int d^2r
dx_adx_b t_A({\bf r})t_B(|{\bf r}-{\bf b}|) \nonumber \\
&&\hspace{-0.1in}\times\
f_{a/A}(x_a,\mu^2,{\bf r}) f_{b/B}(x_b,\mu^2,|{\bf r}-{\bf b}|) \nonumber \\
&&\hspace{-0.1in}\times \frac{d\sigma}{d\hat{t}}(ab\rightarrow cd)\nonumber \\
&&\hspace{-0.1in}\times
\frac{D_{h/c}(z_{c},\mu^2,E, {\bf b}, {\bf r})}{\pi z_{c}} +\mathcal {O}(\alpha_s^3),
\label{eq:AA}
\end{eqnarray}
at fixed impact parameter $\bf b$ in the transverse plane of the beam direction. In Eq.~(\ref{eq:AA})
the average over the
azimuthal angle of the initial fast parton is implicitly implied. The nuclear thickness function $t({\bf r})$
is calculated with the Woods-Saxon distribution function for nucleons in a nucleus and has been normalized by requiring $\int d^{2}r t_{A}({\bf
r})=A$. The nuclear parton
distributions per nucleon (nPDFs) $f_{a/A}(x_a,\mu^2,{\bf r})$ can be parameterized as
the production of the parton distributions inside free nucleons $f_{a/N}(x,\mu^2)$
and the nuclear shadowing factor $S_{a/A}(x,\mu^2,{\bf r})$,
\begin{eqnarray}
   f_{a/A}\left(x,\mu^2,\mathbf{r}\right)&=&S_{a/A}\left(x,\mu^2,\mathbf{r}\right)
  \left[\frac{Z}{A}f_{a/p}\left(x,\mu^2\right)\right.
 \nonumber \\
&~& \left.+\left(1-\frac{Z}{A}\right)f_{a/n}\left(x,\mu^2\right)
\right],\label{eqn:shad}
\end{eqnarray}
where $Z$ denotes the charge and $A$ is the mass number of the nucleus. In the numerical simulations we use the CTEQ6M parametrization
\cite{distribution}
for nucleon parton distributions $f_{a/N}(x,\mu^2)$, and EPS09 parametrization of nPDFs~\cite{EPS09}.
Since the parton-parton scattering cross sections are computed up to NLO, the CTEQ6M parametrization and EPS09 parametrization are both used at
NLO. For simplicity, we only use the central-fit set of EPS09 parametrization in following numerical calculations.

An energetic parton jet produced in the hard scattering may suffer multiple scattering with thermal partons in the QGP
created in nucleus-nucleus collisions. The jet-medium scattering and medium induced gluon radiation should give rise to new contributions to
parton fragmentation functions (FFs) in vacuum and thus leads to medium-modified fragmention functions, which may evolve with the scale $Q$
~\cite{Wang:2003mm,Zhang:2007ja,Zhang:2009rn,Zhang:2008fh,Kang:2014xsa}, in a similar way like the DGLAP evolution in vacuum.
If the parton jet travels a distance $L$ inside the medium with the inelastic scattering mean free path $\lambda$, the probability for the jet
scattering $n$ times to the medium can be assumed to obey Poisson distribution~\cite{Wang:1996yh,Wang:1996pe}. Therefore,
effect of parton energy loss in the dense QCD medium can be calculated  in the high-twist approach of
jet quenching and  the effective medium-modified
parton fragmentation functions (mFFs) can be given by ~\cite{Wang:2003mm,Zhang:2007ja,Zhang:2009rn,Zhang:2008fh},
\begin{eqnarray}
D_{h/c}(z_c,\mu^2,\Delta E_c) = (1-e^{-\langle \frac{L}
{\lambda}\rangle}) \left[ \frac{z_c^\prime}{z_c}
D^0_{h/c}(z_c^\prime,\mu^2) \right.
 \nonumber \\
 \left. + \langle \frac{L}{\lambda}\rangle
\frac{z_g^\prime}{z_c} D^0_{h/g}(z_g^\prime,\mu^2)\right]
+e^{-\langle\frac{L}{\lambda}\rangle} D^0_{h/c}(z_c,\mu^2),
\label{eq:modfrag}
\end{eqnarray}
where $z_c^\prime=p_T/(p_{Tc}-\Delta E_c)$ is the rescaled momentum
fraction of the hadron from the fragmentation of the quenched parton which has the initial transverse momentum $p_{Tc}$ and loses energy $\Delta
E_c$ during its propagation inside the hot medium, $z_g^\prime=\langle L/\lambda\rangle p_T/\Delta E_c$ is the rescaled momentum
fraction of the hadron from the fragmentation of a radiated gluon with initial energy $\Delta E_c/{\langle\frac{L}{\lambda}\rangle}$, and
$z_c=p_T/p_{Tc}$ is the momentum
fraction for jet fragmentation in vacuum. ${\langle\frac{L}{\lambda}\rangle}$ times scattering will provide ${\langle\frac{L}{\lambda}\rangle}$
gluon emissions, so there is a factor ${\langle\frac{L}{\lambda}\rangle}$ in the fragmentation contribution of emitted gluon in above equation.
As shown in Ref. \cite{Wang:1996yh}, the
above mFFs satisfy the momentum sum rule by construction, $\Sigma_h\int zD_{h/c}(z_c,\mu^2,\Delta E_c)=1$.

The weight factor ${\rm exp}({-\langle\frac{L}{\lambda}\rangle})$ is the probability for those partons escaping the medium without suffering any
inelastic scattering, and the weight factor $1-{\rm exp}({-\langle\frac{L}{\lambda}\rangle})$ is the probability for partons encountering at
least one inelastic scattering. The rescaled fraction in Eq. (\ref{eq:modfrag}) is got by energy shifting due to energy loss $\Delta E_c$. For a
given jet, the energy loss $\Delta E_c$ and the scattering number ${\langle\frac{L}{\lambda}\rangle}$ both depend on the local medium density in
the jet trajectory and characterize the medium properties. This approximative approach for medium-modified fragmentation function reproduces the
main effect of medium-induced radiation~\cite{Wang:2003mm}, and is therefore similar to another approximative approach in
references~\cite{BDMS2001, Salgado2003} where the modified fragmentation function is concluded as a convolution of the vacuum fragmentation
function and the probability for a given jet to be quenched to a final jet inside the medium.
We also note the difference between our approach and the ones used in references~\cite{BDMS2001, Salgado2003}.
In higher-twist formalism \cite{Wang:2001ifa,Zhang:2003yn,Majumder:2007hx}, one considers twist-4 processes of the splitting of a highly virtual
parton ($\mu >>
\Lambda_{QCD}$) in QCD medium and evaluates the contribution of medium-induced gluon radiation, which gives rise to the effectively modified
parton fragmentation functions and their corresponding (medium-modified) QCD evolution equations with respect to the hard scale $\mu$. This is
different from that in references~\cite{BDMS2001, Salgado2003} where the medium contribution is computed at the medium scale.  Also in
references~\cite{BDMS2001, Salgado2003} a convolution with a Poisson probability for multiple emissions is used for the energy shift to take
into account the fluctuation of energy loss. 

In a high-twist approach the total energy loss $\Delta E = \Delta E_c(E,{\bf b}, {\bf r})$ is related to the jet transport parameter via,
\begin{eqnarray}
\label{eq:eloss}
  \frac{\Delta E}{E} &=&C_{A} \frac{\alpha_{s}}{2\pi} \int dy^{-} \int_{0}^{Q^{2}} \frac{d l_T^2}{l_{T}^{4}}
  \int_\epsilon^{1-\epsilon} d z
[1+(1-z)^2] \nonumber \\
&&\times\hat q_{F}(y) 4 \sin^{2}(x_{L}p^{+}y^{-}/2),
\end{eqnarray}
as shown in Ref. \cite{deng}, in which one can refer for more details.
$y^-$ denotes a jet place in its trajectory, and is the same as normal time $\tau$ used in following calculations.
In the above expression, the LPM effect for the induced gluon emission originates from the destructive interference of two kinds of processes, i.e. the soft-hard and hard-hard processes, which become identical to each other and lead to a cancellation of their contributions when the transverse momentum of the radiated gluon is small and the formation time of radiated gluon is rather large~\cite{Wang:2001ifa,Zhang:2003yn,Majumder:2007hx}. This is exactly the same as the LPM effect in the double scattering in the GLV opacity expansion formalism~\cite{Gyulassy:2000er4}. In the case of multiple soft scattering, as discussed in references \cite{lpm0,lpm1,Baier:1996sk,Zakharov1996,Ghiglieri:2015zma}
the LPM effect is caused by similar interferences and plays a dominant role when the coherent emission of a single gluon in a multiple scattering process is considered and leads to a suppression of the energy loss as compared to the additive contribution of
$n(=\langle \frac{L}{\lambda}\rangle)$ independent scattering of one gluon radiation.

We emphasize that in our model, the parton matrix elements are calculated at NLO, and EPS nuclear PDFs at NLO are utilized to include the
initial-state cold nuclear matter effects, whereas the parton energy loss due to the final-state hot medium effect is given by a LO derivation
of parton energy loss resulting from effective medium-modified fragmentation functions in twist-4 at the high-twist expansion approach.
Recently several theoretical attempts~\cite{Kang:2013raa,Fickinger:2013xwa, Teaney2013, Wu:2014nca,Ghiglieri:2015zma} have been made to
calculate momentum broadening and parton energy loss due to multiple scattering in medium beyond leading order, which may be incorporated
phenomenologically to a complete NLO calculations of particle or jet productions in high-energy nuclear collisions.  It is also noticed that
recently theoretical investigation of color decoherence has been developed for jets resolving and energy redistributing in the QCD
medium~\cite{Tani2011, Tani2013, Tani2015}.

The jet transport parameter for a gluon is $9/4$ times of a quark,
which is assumed to be proportional to the local parton density in a dynamical evolving medium
and expressed as \cite{Chen:2010te,Chen:2011vt,Casalderrey:2007sw},
\begin{equation}
\label{q-hat-qgph}
\hat{q} (\tau,r)= \hat{q}_0\frac{\rho_{g}(\tau,{\bf r}+{\bf n}\tau)}
{\rho_{g}(\tau_{0},0)}
    \frac{p^\mu u_\mu}{p_0}\,,
\end{equation}
for a parton produced at a transverse position ${\bf r}$ at an initial
time $\tau_0$ and traveling along the direction ${\bf n}$.
$\hat q_{0}$ denotes the jet transport
parameter at the center of the bulk medium at the
initial time $\tau_{0}$.
$\rho_{g}$ is the gluon density at a given temperature $T(\tau, r)$,
and in numerical calculations we assume $\rho_g\propto (1-f)T^3$ for the medium as an ideal gas.
As introduced in Ref.\cite{Chen:2010te,Hirano2001,HT2002}, the fraction $f(\tau, r)$ of the hadronic phase
at any given time and local position is given by
\begin{equation}
f(\tau, r) = \left\{
\begin{array}{l}
0 \\
0 \sim 1 \\
1
\end{array}
 \;\; \;\;
\begin{array}{l}
 \text{if T$>$170 MeV},\\
 \text{if T=170 MeV},\\
 \text{if T$<$170 MeV}
\end{array}
\right.
\end{equation}
where we consider the mixed phase contribution and neglect the pure hadron phase contribution.
During the mixed phase at T = 170 MeV the hadron phase fraction will be $f = 0 \sim 1$ while the QGP phase faction will be $1-f$. With the time
evolution during the mixed phase the fraction value $f$ will increase from 0 to 1. In the following numerical calculations the time-dependent
fraction $f$ is given by simulations of the hydrodynamic model \cite{Hirano2001,HT2002}.
$p^\mu$ is the four momentum of the jet and $u^\mu$ is the
four flow velocity in the collision frame.
The average number of scatterings along the parton propagating path is given by,
\begin{equation}
\label{lamd}
\langle L/\lambda\rangle =\frac{1}{\lambda_0}\int_{\tau_0}^{\infty}
d\tau \frac{\rho_{g}(\tau,{\bf r}+{\bf n}\tau)}
{\rho_{g}(\tau_{0},0)},
\end{equation}
where $\lambda_0$ is the mean free path at the initial time $\tau_{0}$, and for a quark jet it is $9/4$ times of that for a gluon jet.
The parameter $\lambda_0$ as well as $\hat q_{0}$ will be independently inputted
in the following numerical calculations and simultaneously constrained by experiment data.

\begin{figure}
\begin{center}
\includegraphics[width=900mm ]{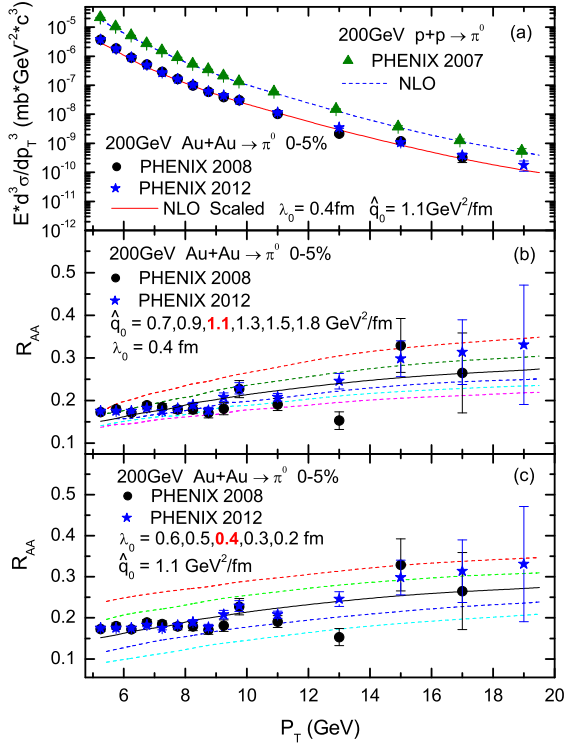}
\caption{(Color online)  (a) The hadron cross sections at mid-rapidity in $p+p$ and central $Au+Au$ collisions at $\sqrt{s_{NN}}=200$ GeV.
(b)(c) The corresponding nuclear modification factor with different values of the jet transport parameter $\hat q_{0}$ and the mean free path
$\lambda_0$. The data are from \cite{Adare:2007dg,PHENIX:2008,PHENIX:2012}.}
\label{fig:RHIC-raa}
\end{center}
\end{figure}
\begin{figure}
\begin{center}
\includegraphics[width=85mm ]{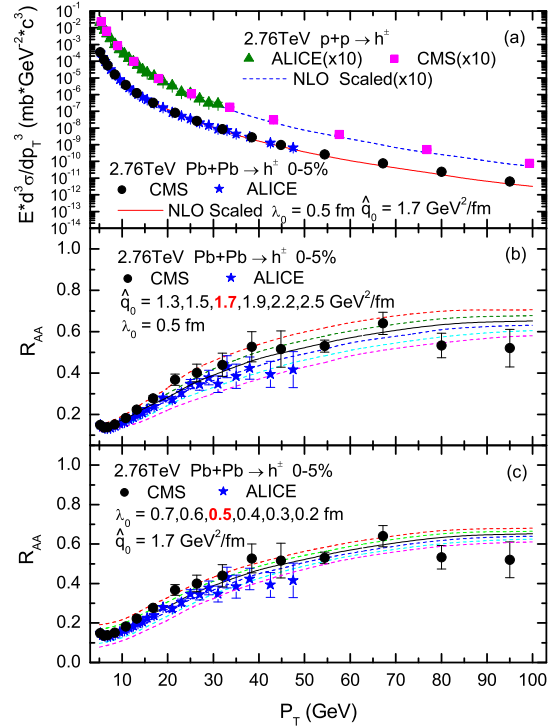}
\caption{(Color online)  (a) The hadron cross sections at mid-rapidity in $p+p$ and central $Pb+Pb$ collisions at $\sqrt{s_{NN}}=2.76$ TeV.
(b)(c) The corresponding nuclear modification factor with different values of the jet transport parameter $\hat q_{0}$ and the mean free path
$\lambda_0$. The data are from \cite{CMS:2012aa,Abelev:2012hxa}.}
\label{fig:LHC-raa}
\end{center}
\end{figure}

The fragmentation function in vacuum $D^0_{h/c}(z_c,\mu^2)$ in Eq.~(\ref{eq:pp}) and (\ref{eq:modfrag}) is given by the AKK parametrization
\cite{akk08}. Here we use the NLO AKK FFs parametrization. For inclusive hadron production given by Eq.~(\ref{eq:pp}) and (\ref{eq:AA}), there
are three independent scales: the renormalization scale $\mu_{ren}$, the factorization scale $\mu_{fact}$ and the fragmentation scale
$\mu_{frag}$. We choose $\mu_{ren}=\mu_{fact}=\mu_{frag}=1.2p_T$ in our numerical analysis for both $p+p$ and $A+A$ collisions.

\section{Extracting parameter values in simultaneous $\chi^2/{\rm d.o.f}$ fits to $R_{AA}$ data}\label{sec:result}

In the model for the effective medium-modified fragmentation functions (mFFs) disucssed in Sec.~\ref{sec:result},
information on the space-time evolution of the local temperature and flow velocity
in the bulk medium along the jet propagation path should be provided. In our simulation we will utilize a (3+1) dimensional ideal
hydrodynamic model \cite{Hirano2001,HT2002} to obtain the space-time evolution of the bulk matter created in central nucleus-nucleus collisions.

With a given space-time profile of the gluon density,
one can then utilize the preceding effective mFFs to obtain the high $p_T$
hadron spectra. In actual calculations for the spectra or cross section at fixed values of the hadron transverse momentum $p_T$ in Eq.
(\ref{eq:pp}) and (\ref{eq:AA}), the factorization and renormalization scales are all chosen as $\mu_f=\mu_R$ = 1.2 $p_T$.
Shown in Fig.~\ref{fig:RHIC-raa} (a) are the hadron cross sections in $p+p$ and 0-5\% $Au+Au$ collisions at $\sqrt{s_{NN}}=200$ GeV with given
parameter values $\hat q_{0} = 1.1$ GeV$^2$/fm and $\lambda_0 = 0.4$ fm, as compared to PHENIX data \cite{Adare:2007dg,PHENIX:2008,PHENIX:2012}.
The theoretical cross section for $A+A$ collisions is scaled as $(dN_{AA}/dyd^2p_T)/{T_{AA}(b)}$ where $T_{AA}(b)=\int d^{2}r t_{A}({\bf
r})t_{A}({\bf r}-{\bf b})$. We choose $b = 2$ fm for 0-5\% $Au+Au$ and $b = 2.1$ fm for 0-5\% $Pb+Pb$ collisions. Shown in
Fig.~\ref{fig:LHC-raa} (a) are for 0-5\% $Pb+Pb$ collisions with given parameter values $\hat q_{0} = 1.7$ GeV$^2$/fm and $\lambda_0 = 0.5$ fm,
as compared to CMS and ALICE data \cite{CMS:2012aa,Abelev:2012hxa}. It is observed that the theoretical results with chosen parameters of $\hat
q_{0}$ and  $\lambda_0$ could describe the experimental data at the RHIC and the LHC very well.

Shown in (b)(c) panels of Fig.~\ref{fig:RHIC-raa} and Fig.~\ref{fig:LHC-raa} are
the suppression factor or nuclear modification factor,
\begin{eqnarray}
R_{AA}=\frac{dN_{AA}/dyd^2p_T}{T_{AA}(b)
d\sigma_{pp}/dyd^2p_T}. \label{eq:raa}
\end{eqnarray}
To compare theoretical results with the RHIC and LHC data, we may fix one parameter of $\hat q_{0}$ or $\lambda_0$, and then choose different
values for another parameter. For
given $\lambda_0$ the nuclear modification factor decreases with the increasing of $\hat q_0$, while for given $\hat q_0$ the nuclear
modification factor increases with the increasing of $\lambda_0$.

For the two parameters $\hat q_0$ and $\lambda_{0}$ we fix one parameter and constrain another by $\chi^2/{\rm d.o.f}$ fitting to data for the
nuclear suppression factor. The $\chi^{2} /{\rm d.o.f}$ is
defined as follows,
\begin{eqnarray}
\chi^2/{\rm d.o.f}=\sum_{i=1}^{N}\Big[\dfrac{(V_{th}-V_{exp})^2}{\sum_t \sigma_t^2}\Big]_i \Big/N,
\end{eqnarray}
where $V_{th}$ stands for the theoretical value, $V_{exp}$ denotes the experimental value, $ \sum_t \sigma_t^2 $
gives the quadratic sum over all types of errors that one chosen point has, and N the number
of data points selected.

\begin{figure}
\begin{center}
\includegraphics[width=90mm ]{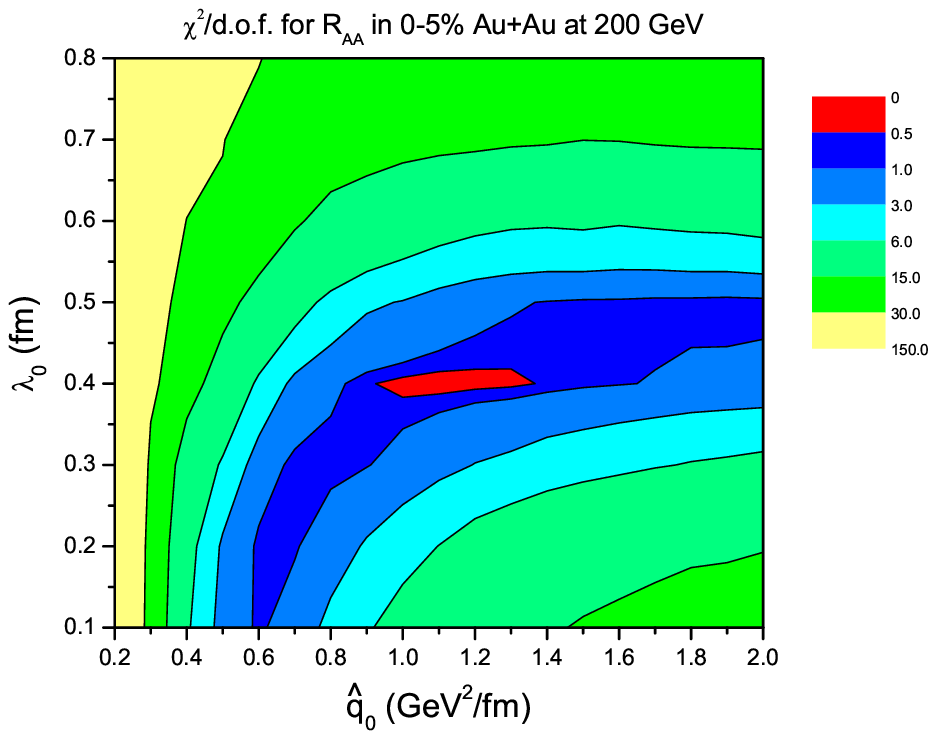}
\includegraphics[width=90mm ]{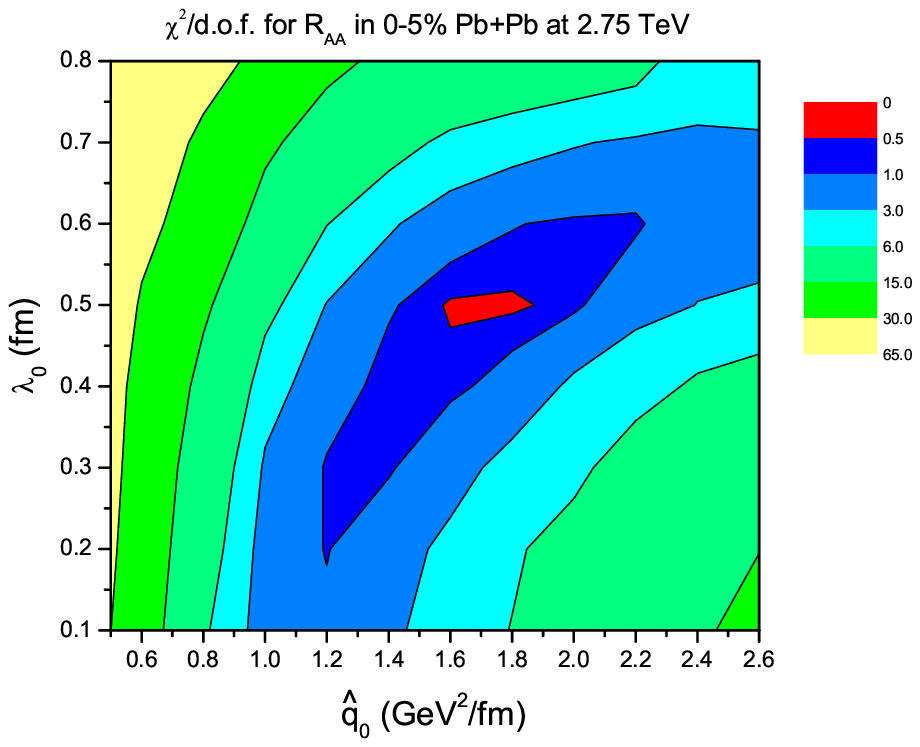}
\caption{(Color online) The $\chi^2/{\rm d.o.f.}$ as a function of the initial quark jet transport parameter $\hat q_0$ and the initial mean
free path $\lambda_{0}$. Upper panel is from fitting to the combined PHENIX data
\cite{Adare:2007dg,PHENIX:2008,PHENIX:2012} on $R_{AA}(p_T)$ for $\pi^0$ with $p_T$ = 5 - 20 GeV at mid-rapidity in $0-5\%$ central $Au+Au$
collisions at $\sqrt{s_{NN}}=200$ GeV.
Lower panel is from fitting to the combined ALICE \cite{Abelev:2012hxa} and CMS \cite{CMS:2012aa} data on $R_{AA}(p_T)$ for charged hadrons with
$p_T$ = 10 - 100 GeV at mid-rapidity in $0-5\%$ central $Pb+Pb$ collisions at  $\sqrt{s_{NN}}=2.76$ TeV.}
\label{fig:RHIC-LHC-chi2}
\end{center}
\end{figure}

In numerical calculations the jet transport parameter $\hat q_0$ and the mean free path $\lambda_0$ are two independent inputs. We choose for a
quark
jet $\hat q_{0}$ = 0.1 - 3.0 GeV$^2$/fm and $\lambda_0$  = 0.1 - 1.0 fm, while for a gluon jet the values are correspondingly multiplied by 9/4
for $\hat q_{0}$ and 4/9 for $\lambda_{0}$ because of different color factors in gluon-gluon  and quark-gluon interacting vertex. From
simultaneous $\chi^2/{\rm d.o.f}$ fits to experimental data
at RHIC and LHC shown in Fig.~\ref{fig:RHIC-LHC-chi2}, one can extract values of the jet transport parameter $\hat q_0$ and the mean free
path $\lambda_0$ at the center of the most central $A+A$ collisions with the given initial time $\tau_0=0.6$ fm.
For a energetic quark jet, the best fits to the combined PHENIX data \cite{Adare:2007dg,PHENIX:2008,PHENIX:2012} give $\hat q_{0}=1.1 \pm 0.30 $
GeV$^2$/fm and $\lambda_0 = 0.4 \pm 0.03$ fm in 0-5\% central $Au+Au$ collisions at $\sqrt{s_{NN}}=200$ GeV,
while the best fits to the combined ALICE \cite{Abelev:2012hxa} and CMS \cite{CMS:2012aa} data lead to
$\hat q_{0}\approx 1.7 \pm 0.3$ GeV$^2$/fm, and $\lambda_{0}\approx 0.5 \pm 0.05$
in 0-5\% central Pb+Pb collisions at $\sqrt{s_{NN}}=2.76$ TeV.

\section{discussions}\label{sec:discussion}

In general, the jet transport parameter should depend on the scale as shown by recent studies on the renormalization of
the jet transport parameter~\cite{Liou:2013qya,Iancu:2014kga,Blaizot:2014bha}.  In our numerical simulations for the initial values for $\hat
q_0$ and $\lambda_0$ given by Eq.~(\ref{q-hat-qgph}) and Eq.~(\ref{lamd}), we assume they
are constants for different jet transverse momentums as a reasonable approximation for phenomenological studies at the RHIC and the LHC. Note
that the best fits shown in Fig.~\ref{fig:RHIC-LHC-chi2} are obtained for hadrons with
$p_T$ = 5-20 GeV at the RHIC while $p_T$ = 10-100 GeV at the LHC, so what we constrain for $\hat q_0$ and $\lambda_0$ may be understood as the
averaged values
for jets with different transverse momentum. Roughly speaking, what we constrain for $\hat q_0$ and $\lambda_0$ should depend on much
greater transverse momentum at the LHC than at the RHIC because of the much wider kinematical region of jet $p_T$ at the LHC.

In the formulation for the medium-modified fragmentation functions in Eq.~(\ref{eq:modfrag}), the final-state medium effect of jet quenching is
controlled both by the total energy loss
$\Delta E \propto \hat q_0$ in Eq.~(\ref{eq:eloss}) and the multiple scattering number $\langle \frac{L}{\lambda}\rangle \propto
\frac{1}{\lambda_0}$ in Eq.~(\ref{lamd}). Therefore the suppression factor $R_{AA}$ is quantified by the two independent parameters $\hat
q_0$ and $\lambda_0$. The trend for the simultaneous $\chi^2/{\rm d.o.f}$ fits in Fig.~\ref{fig:RHIC-LHC-chi2} shows that an increasing $\hat
q_0$ must associate with an increasing $\lambda_0$ to fit well the data at both the RHIC and the LHC. In fact, a larger $\lambda_0$ gives a
smaller scattering
number, and then a larger $\hat q_0$ is needed to release greater energy loss per scattering in order to describe experimental data. This
implicit relation between the two
parameters is consistent with theoretical estimates for the jet transport parameter and the mean free path which are related each other via the
local temperature in a weakly-coupled QCD medium~\cite{Baier:1996sk,Xu:2014ica}.

Of interest are the two different limits as demonstrated in Fig.~\ref{fig:RHIC-LHC-chi2}, the single scattering limit with large $\lambda_0$,
e.g. $\lambda_0$ = 0.7
fm at RHIC, and the infinite number scattering limit with very small $\lambda_0$, e.g $\lambda_0$ = 0.1 fm at LHC.
The numerical simulations for the simultaneous $\chi^2/{\rm
d.o.f}$ fits at both RHIC and LHC show that in the single scattering limit the suppression factor $R_{AA}$ is insensitive to $\hat q_0$ and
sensitive to $\lambda_0$, whereas in the infinite number scattering limit $R_{AA}$ is sensitive to $\hat q_0$ and insensitive to $\lambda_0$.
According to an assumption \cite{Wang:1996yh,Wang:1996pe} for parton scattering obeying a Poisson distribution,
the probability for those partons escaping the system without suffering any inelastic
scattering is ${\rm exp}({-\langle\frac{L}{\lambda}\rangle})$, while the probability for partons encountering at least one inelastic
scattering gives $1-{\rm exp}({-\langle\frac{L}{\lambda}\rangle})$. One can see these two weight factors in
the medium-modified fragmentation function in Eq.~(\ref{eq:modfrag}).
In the infinite number scattering limit with small $\lambda_0$, ${\rm exp}({-\langle\frac{L}{\lambda}\rangle})$ is very small with large
$\frac{L}{\lambda} $, the first term
of Eq. (\ref{eq:modfrag}) with dependence on $\hat q_0$ will dominate the total fragmentation contribution, so $R_{AA}$ is sensitive to $\hat
q_0$ and
less insensitive to $\lambda_0$. On the other hand, in single scattering limit with large $\lambda_0$, the second term of Eq. (\ref{eq:modfrag})
gives the dominant contribution, thus $R_{AA}$ is insensitive to $\hat q_0$ and more sensitive to $\lambda_0$.
We observe that our best fits for $\hat q_0$ and $\lambda_0$ are found in the region between the single scattering limit and the infinite number
scattering
limit due to competing effect between the energy loss per scattering quantified by $\hat q_0$ and the scattering number quantified by
$\lambda_0$, which implies that the data favor a regime of mean-free-paths that suggests multiple scattering in
the medium.

\begin{figure}
\begin{center}
\includegraphics[width=80mm ]{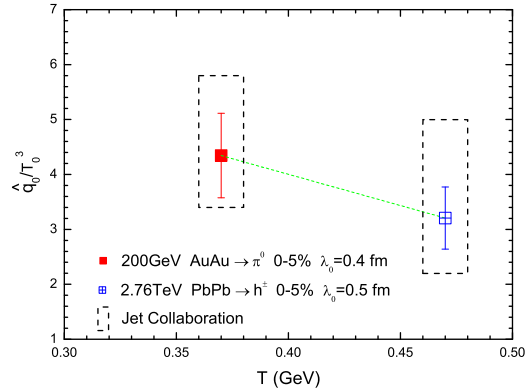}
\caption{(Color online) The scaled jet transport parameter $\hat q/T^3$ for an initial quark jet at the center of the most central A+A
collisions at an initial time $\tau_0=0.6$ fm extracted by comparing the theoretical simulations with experimental data at both RHIC and LHC.
The dashed boxes indicate the corresponding results of Jet Collaboration \cite{Burke:2013yra}.}
\label{fig:q-T0}
\end{center}
\end{figure}

From Fig.~\ref{fig:RHIC-LHC-chi2} we can extract $\hat q_0$ range of values for energetic quarks as constrained by the measured suppression
factors of single hadron spectra at RHIC and LHC as
follows:
\begin{equation*}
\hat q_0 \approx \left\{
\begin{array}{l}
1.1 \pm 0.2  \\
1.7 \pm 0.3
\end{array}
 \;\; {\rm GeV}^2/{\rm fm} \;\; \text{at} \;\;
\begin{array}{l}
 \text{T=373 MeV},\\
 \text{T=473 MeV},
\end{array}
\right.
\end{equation*}
at the highest temperatures reached in the most central Au+Au collisions at RHIC and Pb+Pb collisions at LHC.
As shown in Fig.~\ref{fig:q-T0} for the scaled jet transport parameter $\hat q/T^3$, our result falls within the range of $\hat q_0/T^3_0$ for
energetic quarks extracted from experimental data on $R_{AA}$ by Jet Collaboration
though it is considerably smaller than that given by a strong-coupled AdS/CFT calculation~\cite{Liu:2006ug,Zhang:2012jd} as well as a lattice
calculation~\cite{Majumder:2012sh,Panero:2013pla}.

In addition, from Fig.~\ref{fig:RHIC-LHC-chi2} one can extract $\lambda_{0}$ range of values
for energetic quarks as constrained by the
measured suppression factors of single hadron spectra at RHIC and LHC as:
\begin{equation*}
\lambda_{0} \approx \left\{
\begin{array}{l}
0.4 \pm 0.03  \\
0.5 \pm 0.05
\end{array}
 \;\; {\rm fm} \;\; \text{at} \;\;
\begin{array}{l}
 \text{T=373 MeV},\\
 \text{T=473 MeV}.
\end{array}
\right.
\end{equation*}
In a theoretical estimate \cite{Xu:2014ica} for the mean free path, ${1}/{\lambda_{g}}=\rho\sigma=3\alpha_s(Q^2)T$ for which the elastic cross
section $\sigma$ is used at leading order and the density $\rho$ is for an ideal gas. One can introduce $K$ factor to account for higher order
correction and the more realistic interaction among the medium particles~\cite{Smilga:1996cm,Peigne:2008wu},
\begin{eqnarray}
\frac{1}{\lambda_{g}}=3K\alpha_s(Q^2)T.
\label{eq:lam-g}
\end{eqnarray}
Considering $\lambda_{q}=\frac{9}{4}\lambda_{g}$, and assuming the scale $Q^2=ET$ for a hard parton with energy $E$ traversing a hot QCD medium
with temperature $T$, we find with $K$ = 2.5 - 4.0
$\lambda_{q}$ given by above equation is equal to our best fit for the mean free path at the highest temperatures in both RHIC and LHC.
The $K$ factor is bigger than what would be naturally expected, which might be caused by LO $\sigma$ as well as the ideal gas density $\rho$ in
the theoretical evaluation of ${1}/{\lambda_{g}}=\rho\sigma$. Higher order correction for the cross section may provide a factor of $\sim$ 2,
while other effects such as corrections due to the difference between the real dynamics of the QGP and the simple picture of an ideal gas may
account for the remaining enhancement to $K$. For instance, a strongly interacting QGP may give a larger cross section than a weakly-coupled
QGP. Thus the comparison of the model simulation with the data seems imply that the hot QCD medium at the RHIC and LHC is more likely a strongly
interacting medium, which is surely model-dependent and further validations from other observables will be needed for a robust conclusion.
It is noted that in numerical estimates we use parton energy $E$ = 8 - 25 GeV for $p_T$ = 5 - 20 GeV hadron production at RHIC while $E$ = 15 -
120
GeV for $p_T$ = 10- 100 GeV hadron production at LHC, and therefore the running coupling $\alpha_s(Q^2)$ is appreciably smaller at LHC than at
RHIC.

\begin{figure}
\begin{center}
\includegraphics[width=80mm ]{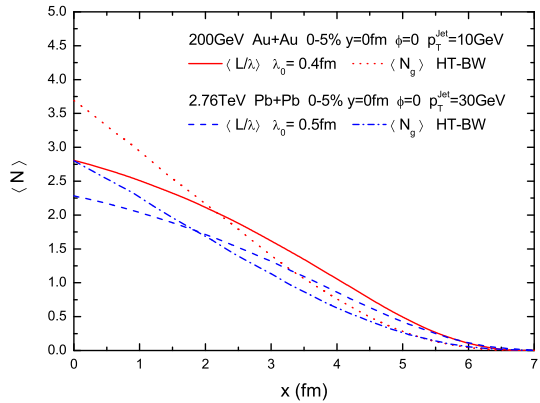}
\caption{(Color online) The averaged number of gluon emissions from a propagating quark as a function of the vertex place of the hard
scattering. The created quarks propagate along $+x$ direction to escape off the fire ball.}
\label{fig:L-lamd-Ng}
\end{center}
\end{figure}

The phenomenological model given by Eq. (\ref{eq:modfrag}) assumes that one scattering will induce one gluon emission from the propagating
parton, so for a given propagating parton the total scattering number equals to the total number of gluon emissions from this parton. Recent
theoretical calculation gives the averaged number of gluon emissions $\langle N_g\rangle$ from a propagating parton in HT-BW approach within the
high-twist framework of parton energy loss \cite{Burke:2013yra, Chang:2014fba}. In the HT-BW model, the medium-modified FFs are given by
numerically solving a set of modified DGLAP evolution equations within the high-twist approach with  an initial
condition given by Poisson convolution of multiple gluon radiations, which has been shown \cite{Chang:2014fba} to give the best agreement with
data for the nuclear modification factor $R_{AA}$ in high-energy heavy-ion collisions. Especially, the averaged number of gluon emissions
$\langle N_g\rangle$ from a propagating parton is given \cite{Chang:2014fba} in the study for modified DGLAP evolution equations, which can be
compared with our extracted number of medium-induced emissions $\langle\frac{L}{\lambda}\rangle$. Shown in Fig.~\ref{fig:L-lamd-Ng} is the
comparison for gluon emission number
between our model (solid and dash curves) and the HT-BW approach (dot and dotted-dash curves denoted as ``HT-BW"), where the initial quark jets
are produced in the point $(x,y=0)$ of $x$ axis and propagate along $+x$ direction in the
transverse plane to escape off the fire ball. The quark transverse momentum is for example chosen as $p_T^{\rm jet}$ = 10 GeV for RHIC and 30
GeV for LHC in central A+A collsions. Our results for the averaged number of gluon emissions are consistent with the HT-BW method, and justify
the validity of the model as shown in Ref.~\cite{Zhang:2007ja,Zhang:2009rn}.

\begin{figure}
\begin{center}
\includegraphics[width=80mm ]{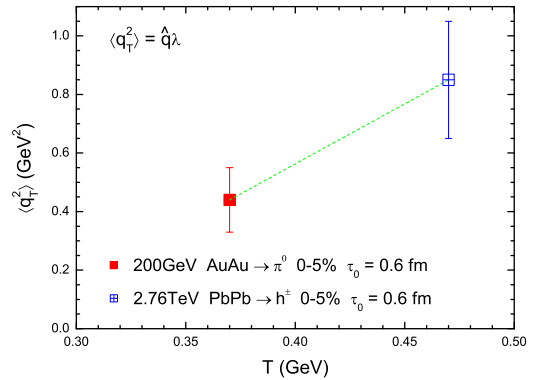}
\caption{(Color online) The temperature dependence of the average transverse momentum broadening squared for energetic quarks for one scattering
at the initial time $\tau_0$ in the center of the fireball.}
\label{fig:q-lam-T0}
\end{center}
\end{figure}

According to definitions for the jet transport parameter and the mean free path \cite{Baier:1996sk},
\begin{equation}
\hat q=\rho \int dq_T^2 \frac{d\sigma}{dq_T^2} q_T^2,
\label{eq:qhat}
\end{equation}
\begin{equation}
\frac{1}{\lambda}=\rho \int dq_T^2 \frac{d\sigma}{dq_T^2},
\label{eq:lamda}
\end{equation}
one can estimate the average transverse momentum broadening squared,
\begin{equation}
\langle q_T^2\rangle=\hat q \lambda.
\label{eq:broaden}
\end{equation}
Then our best fitting values for $\hat q_0$ and $\lambda_{0}$ give,
\begin{equation*}
\langle q_T^2\rangle=\hat q \lambda \approx \left\{
\begin{array}{l}
0.44 \pm 0.11  \\
0.85 \pm 0.20
\end{array}
 \;\; {\rm GeV}^2 \;\; \text{at} \;\;
\begin{array}{l}
 \text{T=373 MeV},\\
 \text{T=473 MeV},
\end{array}
\right.
\end{equation*}
for energetic quarks with one scattering at the initial time $\tau_0$ in the center of the fireball, as shown in Fig.~\ref{fig:q-lam-T0}. Our
numerical results show that for energetic parton jets scattering inside the bulk medium at the highest temperature, the average transverse
momentum broadening squared at LHC is about twice of that at RHIC. Compared to initial parton jet energy, the broadening is moderate, which
implies that the jet may traverse through the medium with small reflection and justifies the eikonal approximation usually used in jet quenching
calculations.

As we stated before the AKK FFs in vacuum are used in our numerical simulations. It is noted that a recent theoretical study \cite{Enterria} has
confronted seven sets of NLO FF parameterizations with inclusive charged-particle spectra in p+p collisions at the LHC and identified that most
of the theoretical predictions including AKK08 tend to overpredict the measured cross sections by up to a factor of two due to the too-hard
gluon-to-hadron FFs. In this paper
we focus on the medium properties demonstrated by the nuclear modification factor $R_{AA}$ which is a ratio of spectra between $A+A$ and $p+p$
collisions and therefore is expected not to be very sensitive to the choice of FFs parametrization as well as the scale.  We have redone our
simulations with Kretzer parametrization of FFs~\cite{kre}, which show the extracted ${\hat q}_0$ ($\lambda_0$) is less (larger) about 10-20\%
by using AKK08 FFs than by using Kretzer FFs.

~

\section{Summary}\label{sec:summary}

We have used the NLO pQCD parton model with effective
modified fragmentation functions due to radiative parton energy
loss to study single hadron spectra in high-energy heavy-ion
collisions at both RHIC and LHC. The energy loss of the hard partons is incorporated in the modified fragmentation functions which utilize two
most important parameters to characterize the properties of the bulk medium, the jet transport parameter $\hat q_{0}$ and the mean free path
$\lambda_{0}$ both at the initial time $\tau_0$.
We perform the phenomenological study of the experimental data for $R_{AA}(p_{T})$ to
constrain the two parameters with simultaneous $\chi^2/{\rm d.o.f}$ fits to RHIC as well as LHC data,
and obtain for energetic quarks $\hat q_{0}\approx 1.1 \pm 0.2$ GeV$^2$/fm and $\lambda_{0}\approx 0.4 \pm 0.03$ fm in central $Au+Au$
collisions at $\sqrt{s_{NN}}=200$ GeV, while $\hat q_{0}\approx 1.7 \pm 0.3$ GeV$^2$/fm, and $\lambda_{0}\approx 0.5 \pm 0.05$ fm in central
$Pb+Pb$ collisions at $\sqrt{s_{NN}}=2.76$ TeV.
Numerical analysis shows that the best fit falls between the single scattering limit and multiple scattering limit for the energetic jets
propagating through the bulk medium.
These results indicate that the average transverse momentum broadening squared $\langle q_T^2 \rangle = \hat q\lambda$ of energetic partons for
one scattering at initial time $\tau_0$ in the center of the fireball at LHC, $\langle q_T^2 \rangle_{LHC} \approx 0.85$ GeV$^2$, which is twice
of $\langle q_T^2 \rangle_{RHIC} \approx 0.44$ GeV$^2$ at RHIC.

~

\section*{Acknowledgements}

The authors thank X.-N. Wang, G.-Y. Qin and N.-B. Chang for stimulating discussions. This work is supported by the Major State Basic Research
Development
Program in China under Contract No. 2014CB845400, and by Natural Science Foundation of China under Project Nos. 11435004, 11175071, 11322546 and
11221504.

\end{document}